\documentclass[12pt,fleqn]{article}
\usepackage{epsfig,rotating}
\begin{document}

\title{\bf Parity Violation in $\vec{p}p$ and $\vec{n}p$ Experiments}

\author{\large W.D. Ramsay\\
\em Department of Physics and Astronomy,\\
\em University of Manitoba, Winnipeg, MB, R3T 2N2, Canada}

\maketitle

\begin{abstract}

Parity violation experiments involving only two nucleons provide a way to
study the non-leptonic, strangeness-conserving part of the weak interaction
in a clean measurement free of nuclear structure uncertainties. Such
measurements are particularly appropriate for discussion at this conference
as their success depends critically on the ability to accurately control
and measure spin. Although simple in principle, the experiments are
technically very demanding and great pains must be taken both in the
preparation of the incident polarized beams and the measurement of the
resultant parity violating asymmetries, which may be masked by a multitude
of systematic effects. At low and intermediate energies, $\vec{p}p$
experiments are sensitive to the medium range part of the parity violating
nucleon-nucleon force, usually parameterized in terms of rho and omega
meson exchange. The pion does not contribute to parity violation in the
$pp$ experiments, as the $\pi^0$ is its own antiparticle and parity
violation would also imply CP violation. I review existing $pp$
measurements with particular emphasis on the recent 221 MeV $\vec{p}p$
measurement at TRIUMF which permitted the weak meson-nucleon coupling
constants $h^{pp}_\rho$ and $h^{pp}_\omega$ to be determined separately for
the first time. The $\vec{n}p$ experiments, on the other hand, are used to
extract the weak pion nucleon coupling, $f_\pi$, describing the longest
range part of the parity violating nucleon-nucleon force. The $np$ system
is the only two nucleon system that is sensitive to $f_\pi$. I also review
these experiments, with specific details of the $\vec{n}p \rightarrow
d\gamma$ experiment now under preparation at Los Alamos National
Laboratory.

\end{abstract}

\section{Introduction}

\begin{quote}
{\em Sometimes is is necessary to repeat what we know.  All mapmakers
should place the Mississippi at the same location, and avoid originality.
--} Saul Bellow
\end{quote}

In preparing a review talk, one becomes acutely aware of essentially
``repeating what we know''.  In this review I will, however, do just that,
concentrating in particular on what we know about $pp$ and $np$ parity
violation experiments in which we control the spin of the incident nucleon,
in other words experiments that use the nucleon spin as a tool, rather than
experiments concerned with the nature of the nucleon spin itself.  I will
give a historical overview of $\vec{p}p$ and $\vec{n}p$ parity violation
experiments, with technical details of the experiments I am most familiar
with -- the TRIUMF 221 MeV $\vec{p}p$ experiment and the Los Alamos
$\vec{n}p \rightarrow d\gamma$ experiment now being installed at LANSCE.
What I cover is a  biased personal selection, and I refer readers
interested in more background to two fine reviews of the field by
Adelberger and Haxton \cite{Ade85} and Haeberli and Holstein \cite{Hae95}.

\begin{figure}
\begin{center}
\includegraphics[height=0.8\textwidth,angle=-90]{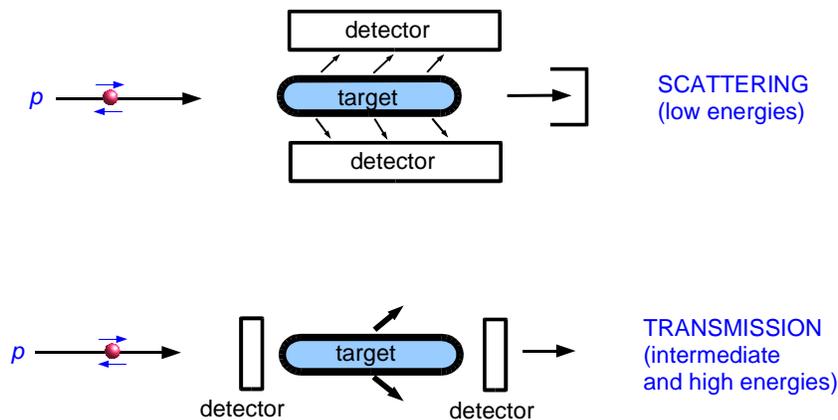}
\end{center}
\caption{Types of $\vec{p}p$ experiments. The low-energy experiments use
scattering geometry, while the intermediate and high-energy experiments
use transmission geometry.}
\label{pptypes}
\end{figure}

\subsection{Parity and the Weak Interaction}
The parity operation reflects all space coordinates through the origin
($\vec{r} \rightarrow -\vec{r}$), which is equivalent to a mirror
reflection (which reverses only one space coordinate)\footnote{The answer
to your kids' question, ``Why does a mirror reverse left-right but not
up-down?'' is that it doesn't; it reverses fore and aft. You do the
left-right reversal when you turn the paper around to look at it in the
mirror.} and a 180$^0$ rotation (which reverses the other two). Since we
can assume rotational invariance, the parity operation is often thought of
as simply a mirror reflection.  If a process is not identical to its mirror
image, it is said to be parity violating (PV) or parity nonconserving
(PNC). In physics, parity violation is exclusive to processes involving the
weak interaction, such as:

\begin{enumerate}
\item $\mu \rightarrow e^- + \nu_{\mu} \overline{\nu}_e$
\item $n \rightarrow p + e^- + \overline{\nu}_e$;~~~~~~~$\Lambda
\rightarrow p + e^- + \overline{\nu}_e$
\item $K^+ \rightarrow \pi^+ \pi^-$
\item $pp \rightarrow pp$
\end{enumerate}

The first three examples, in the leptonic, semi-leptonic and hadronic
$\Delta s = 1$, sectors, clearly involve the weak interaction; the decays
shown would disappear without it.  The last example, however, in the purely
hadronic $\Delta s = 0$ sector, is so dominated by the strong interaction
that, were the weak interaction to disappear, the process would be
virtually unchanged. The only way to see the effects of the weak interaction
in such processes is to look for parity violation, and the
experimental signal is very small.  In $pp$ scattering, for example, the
parity violating part of the scattering cross section is typically of order
$10^{-7}$ of the parity conserving part.

\begin{table}[!b]
{\scriptsize
\begin{tabular}{llcr}
\hline\hline
  \multicolumn{1}{c}{\bf Lab/Energy} &
  \multicolumn{1}{c}{\bf Technical Details} &
  \multicolumn{1}{c}{\bf $A_z$ ($10^{-7}$)} &
  \multicolumn{1}{c}{\bf Where Reported}\\
\hline
Los Alamos & scattering                  & $+1 \pm 4$       & 1974 Phys. Rev. Lett. \cite{Pot74}     \\
15 MeV     & 3 atm x 38cm hydrogen gas   &                  &                                        \\
           & 4 liquid scintillators      &                  &                                        \\
\hline
           & scattering                  &  $-1.7 \pm 0.8$  & 1978 Argonne Conference \cite{Nag79}   \\
           & 6.9 atm hydrogen gas        &                  &                                        \\
           & 4 plastic scintillators     &                  &                                        \\
\hline
Texas A\&M & scattering                  & $-4.6 \pm 2.6$   & 1983 Florence Conference \cite{Tan83}  \\
47 MeV     & 39 atm x 42cm hydrogen gas  &                  &                                        \\
           & 4 plastic scintillators     &                  &                                        \\
\hline
Berkeley   & scattering                  & $-1.3 \pm 1.1$   & 1980 Santa Fe Conference \cite{vonR80} \\
46 MeV     & 80 atm hydrogen gas target  &                  &                                        \\
           & He ion chamber around target& $-1.63 \pm 1.03$ & 1985 Osaka Conference \cite{vonR85}    \\
\hline
SIN (PSI)  & scattering                  & $-3.2 \pm 1.1$   & 1980 Phys. Rev. Lett. \cite{Bal80}     \\
45 MeV     & 100 atm hydrogen gas        &                  &                                        \\
           & annular ion chamber         & $-2.32 \pm 0.89$ & 1984 Phys. Rev. D. \cite{Bal84}        \\
           &                             &                  &                                        \\
           &                             & $-1.50 \pm 0.22$ & 1987 Phys. Rev. Lett. \cite{Kist87}    \\
\hline
Los Alamos & transmission                & $+2.4 \pm 1.1$   & 1986 Phys. Rev. Lett. \cite{Yua86}     \\
800 MeV    & 1 m liquid hydrogen gas     &                  &                                        \\
           & ion chambers                &                  &                                        \\
\hline
Bonn       & scattering                  & $-1.5 \pm 1.1$   & 1991 Phys. Lett. B \cite{Evers91}      \\
13.6 MeV   & 15 atm hydrogen gas         &                  &                                        \\
           & hydrogen ion chambers       & $-0.93 \pm 0.21$ & 1994 private communication \cite{Evers94}\\
\hline
TRIUMF     & transmission                & $+0.84 \pm 0.34$ & 2001 Phys. Rev. Lett. \cite{Berd01b}   \\
221 MeV    & 40 cm liquid hydrogen       &                  &                                        \\
           & hydrogen ion chambers       &                  &                                        \\
\hline
Argonne ZGS& transmission                & $+26.5 \pm 7.0$ & 1986 Phys. Rev. Lett. \cite{Loc84}      \\
5130 MeV   & 81 cm water target          &                  &                                        \\
           & ion chambers                &                  &                                        \\
           & and scintillators           &                  &                                        \\
\hline\hline
\end{tabular}
}
\caption{Summary of $\vec{p}p$ parity violation experiments. The long times
taken to achieve small uncertainties reflects the time taken to understand
and correct for systematic errors. In cases where authors reported both
statistical and systematic uncertainties, this table shows the quadrature
sum of the two.}
\label{ppexp}
\end{table}

\section{$\vec{p}p$ Experiments}

Figure \ref{pptypes} shows typical $\vec{p}p$ parity violation experiments.
They scatter a longitudinally polarized beam of protons from a hydrogen
target and measure the difference in cross section for right-handed and
left-handed proton helicities. The low energy experiments use scattering
geometry, in which the detectors measure the scattered protons directly.
The intermediate and high energy experiments use transmission geometry in
which the change in scattering cross section is deduced from the change in
transmission through the target. Transmission geometry uses a simpler
detector arrangement, but can't be used at low energies because the energy
loss in the target is too high to permit a sufficiently thick target. The
quantity reported by both types of experiments is the parity violating
longitudinal analyzing power, $A_z = \frac{\sigma^+ - \sigma^-}{\sigma^+ +
\sigma^-}$, where $\sigma^+$ and $\sigma^-$ are the scattering cross
sections for positive and negative helicity. Because the statistical
precision required on $A_z$ is typically $\pm 10^{-8}$, it would take too
long to count the requisite $10^{16}$ scattered particles, and all
$\vec{p}p$ experiments so far have used current mode detection (as opposed
to counting individual scattered particles).

\begin{figure}
\begin{center}
\includegraphics[width=0.45\textwidth]{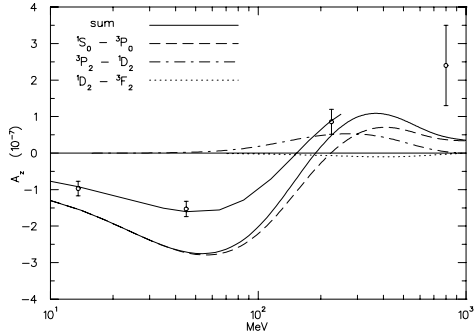}
\end{center}
\caption{Partial wave decomposition of $A_z$. The broken curves and lower
solid sum curve are calculated by Driscoll and Miller \cite{Dris89} using
the DDH  best guess couplings \cite{DDH80}. The upper solid sum curve is
calculated by Carson {\em et al.} \cite{Carl01} with adjusted couplings.}
\label{azpw}
\end{figure}

\subsubsection{Historical Summary}

A roughly historical summary of $\vec{p}p$ parity violation experiments is
given in Table 1. The long time taken to acquire measurements at a
reasonable selection of energies and with small experimental uncertainties
reflects the technical difficulty of these measurements.  Running time is
dominated by the time required to understand, and correct for, the various
sources of systematic error. The time required to get the desired
statistical precision is normally small by comparison.

\subsubsection{Interpretation of the Results}

Although it is known that the weak force is carried by the W and Z bosons,
it is also known that their range is very small ($\sim 0.002 \:fm$), and
most authors assume that at low and intermediate energies the protons
never get close enough for direct W and Z exchange. The interaction is
normally treated in a meson exchange model with one strong, parity
conserving vertex and one weak, parity non-conserving vertex. The weak
vertex is parameterized in terms of a set of weak meson-nucleon coupling
constants, $f_{\pi} , h_{\rho}^{0,1,2} , h_{\omega}^{0,1}$, where the
subscript denotes the exchanged meson and the superscript gives the isospin
change \cite{DDH80}. Mesons heavier than the $\omega$-meson are not
included because of the hard core of the nucleon-nucleon force. Further,
for the $pp$ interaction there is no $\pi$ exchange because the $\pi^{0}$
is its own antiparticle and parity violation would also imply CP violation,
another factor of 10$^3$ suppression. CP invariance also excludes other
neutral scalar and pseudoscalar mesons such as the $\eta$,
$\eta^{\prime}$, $S$, and $\delta^0$ from consideration (Barton's theorem
\cite{Bar61}).

The weak couplings were calculated by Desplanques, Donoghue and Holstein
\cite{DDH80}, and subsequently by a number of other authors
\cite{Dub86,Feld91,Kais89,Meis90,Meis99}. The range of calculated values is
large, and an experimental determination is needed.

\begin{figure}[t]
\begin{center}
\includegraphics[width=0.45\textwidth]{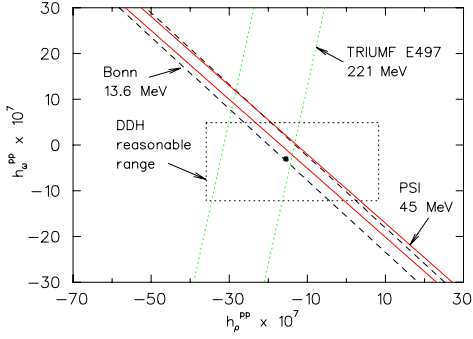}
\end{center}
\caption{Constraints on the weak meson-nucleon couplings imposed by
experiments in the energy range where the meson exchange model is normally
used. The bands are based on calculations by Carlson {\em et al.}
\cite{Carl01} using the AV18 potential \cite{Wirin95} and CD-Bonn strong
couplings \cite{Mach01}. (Figure modified from \cite{Berd01b})}.
\label{limits}
\end{figure}

Figure \ref{azpw} shows $A_z$ as a function of energy, broken down into
contributions from various partial wave mixings \cite{Dris89}. Since the
parity is $(-1)^\ell$, where $\ell$ is the orbital angular momentum, one
would not normally expect partial waves differing by one unit in $\ell$
(S-P, P-D, etc.) \footnote{The notation in Fig. \ref{azpw} is
$^{(2S+1)}{\ell}_J$, where S is the total spin (0 or 1), $\ell$ = S,P,D,F
is orbital angular momentum of 0,1,2,3, and J is the total angular
momentum.} to mix, but due to the weak force, some small mixing occurs. At
low energy, $A_z$ is dominated by the contribution of S-P mixing. At the
energy of the TRIUMF experiment, $A_z$ arises almost exclusively from P-D
mixing. The {\em shape} of the curves is set by the strong interaction,
while the multiplying factor is set by the weak interaction. By adjusting
the multiplying factors to fit the data, the weak meson-nucleon couplings
can be extracted. The lower solid line is the total $A_z$ calculated by
Driscoll and Miller \cite{Dris89} using the DDH \cite{DDH80} weak
couplings. The upper solid line is from a calculation by Carlson {\em et
al.} \cite{Carl01} using adjusted values of the weak couplings.

$pp$ experiments are sensitive to the combinations $h_{\rho}^{pp} =
h_{\rho}^{(0)}+h_{\rho}^{(1)}+ \frac{1}{\sqrt{6}} h_{\rho}^{(2)}$ and
$h_{\omega}^{pp}=h_{\omega}^{(0)}+h_{\omega}^{(1)}$. Using the AV18 strong
potential \cite{Wirin95} and CD-Bonn values for the strong couplings
\cite{Mach01}, Carlson {\em et al.} \cite{Carl01} calculate that
\begin{eqnarray*}
A_z(13.6 MeV) & = &  0.059h_{\rho}^{pp}+0.075h_{\omega}^{pp} \\
A_z(45 MeV)   & = &  0.10h_{\rho}^{pp}+0.14h_{\omega}^{pp} \\
A_z(225 MeV)  & = & -0.038h_{\rho}^{pp}+0.010h_{\omega}^{pp}
\end{eqnarray*}
These constraints are shown graphically in Fig. \ref{limits}. Notice that
the precision results from Bonn and PSI determine essentially the same
combination of couplings.

\begin{table}
\caption{Overall corrections for systematic errors. The table shows
the average value of each coherent modulation, the net correction made for
this modulation, and the uncertainty resulting from applying the
correction.}
\begin{tabular}{lcr}
\hline
 Property & Average Value & $10^7 \Delta A_z$  \\
\hline
 $A_z^{uncorrected} (10^{-7})$ & ~~~~~~~$1.68 \pm 0.29(stat.)$ & \\
 $y*P_x (\mu m)$ & $-0.1 \pm 0.0$ & $-0.01 \pm 0.01$ \\
 $x*P_y (\mu m)$ & $-0.1 \pm 0.0$ & $0.01 \pm 0.03$ \\
 $\langle yP_x \rangle (\mu m)$ & $1.1 \pm 0.4$ & $0.11 \pm 0.01$ \\
 $\langle xP_y \rangle (\mu m)$ & $-2.1 \pm 0.4$ & $0.54 \pm 0.06$ \\
 $\Delta I/I (ppm)$ & $15 \pm 1$ & $0.19 \pm 0.02$ \\
 $position + size$ &           & $     0  \pm 0.10$ \\
 $\Delta E(meV at\, OPPIS)$&   7--15      & $  0.0  \pm 0.12$ \\
 electronic crosstalk &     & $ 0.0 \pm 0.04$ \\
 Total & & $0.84 \pm 0.17 (syst.)$ \\
\hline\hline
 $A_z^{corr} (10^{-7})$ &
\multicolumn{2}{c|}{$0.84 \pm 0.29(stat.) \pm 0.17(syst.) $} \\
 $\chi_{\nu}^2 (23 sets)$ &
\multicolumn{2}{c|}{1.08} \\
\hline
\end{tabular}
\label{corr}
\end{table}

\begin{figure}[t]
\begin{center}
\includegraphics[width=.5\textwidth,height=9cm]{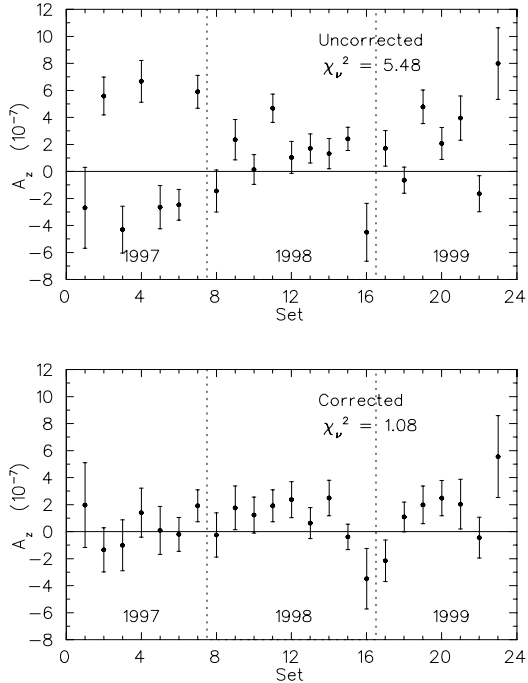}
\end{center}
\caption{Effect of corrections to the E497 data. The main effect visible
to the eye is for first moments of transverse polarization. }
\label{E497corr}
\end{figure}

The TRIUMF $pp$ experiment \cite{Berd01b} is a transmission experiment, as
depicted in the bottom panel of Fig. \ref{pptypes}. A 221 MeV
longitudinally polarized proton beam was passed through a 40 cm long liquid
hydrogen target, which scattered about 4\% of the beam. Hydrogen filled ion
chambers located upstream and downstream of the target measured the change
in transmission when the spin of the incident protons was flipped from
right-handed to left-handed. Although a very good optically pumped
polarized ion source \cite{Zelen96,Zelen97,Lev95} was used that minimized
the changes in beam properties other than helicity, other beam properties
still change very slightly. These helicity-correlated beam property
changes cause a systematic shift in the $A_z$ distribution, and corrections
must be made. To do this, the TRIUMF group continuously measured the
helicity correlated changes in beam properties and made corrections based
on the sensitivities determined in separate control measurements. The data
before and after correction are shown in Fig. \ref{E497corr}. The main
effect visible to the eye is from {\em first moments of transverse
polarization} resulting from the distribution of transverse polarization
components across the beam. All the corrections are summarized in Table 2.
The measured $A_z$ actually came half from true parity violation and half
from false effects.

\section{$\vec{n}p \rightarrow d\gamma$ Experiments}

Unlike the $\vec{p}p$ experiments just discussed, which are sensitive to
$\rho$ and $\omega$ exchange,  $\vec{n}p \rightarrow d\gamma$ experiments
are sensitive almost exclusively to pion exchange, and measure the weak
pion-nucleon coupling, $f_{\pi}$.\footnote{Some authors quote
$H_{\pi}=f_{\pi}\frac{g_{\pi}}{\sqrt{32}}$, where $g_{\pi}$ is the strong
pion-nucleon coupling.} Measurements such as the circular $\gamma$-ray
polarization from $^{21}$Ne \cite{Ear83}, or the longitudinal analyzing
power in $p\alpha$ scattering \cite{Lan86}, provide constraints on a
combination of $\pi$, $\rho$, and $\omega$ couplings. The most precise
limit on $f_{\pi}$ alone is believed to be from measurements of circularly
polarized gamma rays from a parity mixed doublet in $^{18}F$
\cite{Page87a,Bini85}. These results, however, are only about 10\% of
theoretical predictions \cite{DDH80,Kap93,Hen98}, which give $f_{\pi} \sim
4 \times 10^{-7}$, and are also at odds with the large value of $f_{\pi}$
deduced from measurements of the anapole moment in $^{133}Cs$
\cite{Wood97,Flam97}, although the $^{133}Cs$ experiment is also sensitive
to $h^{(0)}_{\rho}$, and, as pointed out by Wilburn and Bowman,
\cite{Wil98} the disagreement may not be too significant.

$np \rightarrow d\gamma$ radiative capture measurements can be made with
unpolarized neutrons, as was done in the Leningrad experiment \cite{Kny84},
but the circular polarization, $P_\gamma$, of the capture gammas must be
measured and the analyzing power of the polarimeters is very low.
In an $\vec{n}p \rightarrow d\gamma$ experiment, the incident cold neutrons
are polarized vertically and the gamma rays produced by neutron capture in
the hydrogen target are expected to be emitted slightly more in the
direction opposite to the neutron spin. The up-down asymmetry $A_{\gamma}
\approx -0.11 f_{\pi}$ provides a clean measure of $f_{\pi}$ free of
nuclear structure uncertainties \cite{Snow00}. Previous measurements at ILL
Grenoble gave $A_{\gamma}=(6\pm 21) \times 10^{-8}$ \cite{Cav77} and
 $A_{\gamma}=(-1.5\pm 4.8) \times 10^{-8}$ \cite{Alb88}, but neither result
was accurate enough to impose a significant constraint.

An experiment is now being prepared at Los Alamos to measure the gamma ray
asymmetry in  $\vec{n}p \rightarrow d\gamma$ with an uncertainty of $\pm
0.5 \times 10^{-8}$ \cite{Snow00}. The expected asymmetry is $A_{\gamma}
\approx -5 \times 10^{-8}$.

\begin{figure}
\begin{center}
\includegraphics[width=\textwidth]{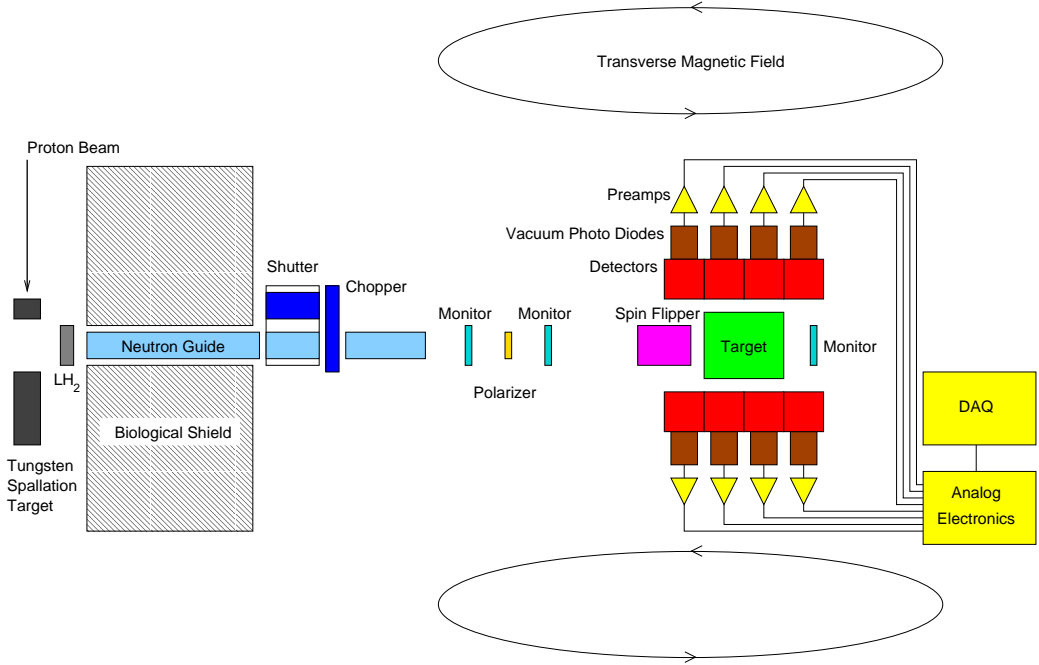}
\end{center}
\caption{Layout of apparatus for the $\vec{n}p \rightarrow d\gamma$
experiment at LANSCE.}
\label{npdgexp}
\end{figure}

The apparatus is shown schematically in Fig. \ref{npdgexp}. A pulsed, 800
MeV, 100-150 $\mu$A proton beam impinges on a tungsten spallation target.
Neutrons from the spallation target are cooled in a liquid hydrogen
moderator and transported to the experiment in a neutron guide. The neutron
guide prevents the $1/r^2$ intensity fall-off that would otherwise occur,
and also enhances the fraction of low energy neutrons. The peak neutron flux
through the 9.5 cm x 9.5 cm guide is $6 \times 10^7$ n/ms at 8 meV.

The neutrons are polarized by passing through a polarized $^3He$ spin
filter. Neutrons with spin parallel to the $^3He$ spin pass through the
filter, while neutrons with spin anti-parallel to the $^3He$ spin are
captured. The neutron polarization is determined by the front monitors
located before and after the spin filter. If half the neutrons are
captured, the polarization of the downstream neutron beam is 100\%. A test
$^3He$ spin filter was demonstrated in a fall 2000 test run and the final
version will be tested at the University of Michigan in fall 2002.

The polarized neutrons are captured in a 30 cm long, 20 L liquid
para-hydrogen target operating at 17 K. The low temperature is important to
keep a small (0.03\%) equilibrium ortho-hydrogen fraction, as the spin 1
ortho-hydrogen molecule allows spin-flip scattering and destroys the
asymmetry. The cross section for scattering from ortho-hydrogen is
approximately 20 times that for scattering from para-hydrogen, so the
neutron detector downstream of the target will indicate any change in the
ortho-hydrogen fraction. Approximately 60\% of the neutrons are captured in
the para-hydrogen, producing a deuteron and a 2.2 MeV gamma ray. The gamma
rays are measured by an array of 48 15 x 15 cm$^2$ CsI(Tl) detectors
surrounding the target.

The neutron spin is flipped 20 times per second by a 30 kHz RF spin filter.
The RF itself has a very small ($<0.1$ mm) skin depth in, and is well
contained by, the conducting shell surrounding the spin filter. In the fall
2000 test runs at LANSCE, the on-axis spin flipper efficiency was >95\%.

The neutron beam is pulsed at 20 Hz, so the data acquisition operates in 50
ms ``frames''. The pulsed beam makes it possible to identify neutrons by
their time of flight over the 22 m flight path. The fast neutrons arrive at
the target first, and for the first 9 ms the neutron energy is above the
15 meV required to excite a para-to-ortho transition in the para-hydrogen
target. For this reason, the first 9 ms of a frame will have no
physics asymmetry and can be used to measure background.  From 9 ms to 30
ms, the neutron energy falls from 15 meV to 1.5 meV. The amplitude of the
RF spin filter is synchronized with this fall to ensure that fast and slow
neutrons are all rotated by 180$^0$. After 40 ms, a ``frame overlap
chopper'' blocks the neutrons so that slow neutrons from one frame are not
still arriving when fast neutrons from the next pulse arrive. During the 40
ms to 50 ms dead interval, electronic noise can be checked. Different
systematic errors have a different dependence on neutron energy and their
time-of-flight signatures can be used to identify them. In addition, the
$^3He$ spin filter direction and the overall holding field can be reversed
for further cancellation of systematic errors.

The beamline, FP12, is now almost complete at LANSCE. The experimental cave
is scheduled for installation in early 2003 and commissioning runs are
planned for summer and fall of 2003.

\section{Summary}

$pp$ and $np$ parity violation experiments provide a means to study the
hadronic, $\Delta s = 0$ part of the weak interaction.  $\vec{n}p
\rightarrow d \gamma$ experiments are sensitive to the long range part of
the nucleon-nucleon interaction and constrain the weak pion-nucleon
coupling constant, $f_\pi$.  Despite decades of experimental and
theoretical work, the strength of this coupling is still very uncertain.
The new, precision experiment now under construction at the LANSCE pulsed
neutron source at Los Alamos should finally lay this question to rest.
$\vec{p}p$ parity violation experiments are sensitive to the shorter range
part of the nucleon-nucleon force and constrain the combinations
$h_{\rho}^{pp} = h_{\rho}^{(0)}+h_{\rho}^{(1)}+ \frac{1}{\sqrt{6}}
h_{\rho}^{(2)}$ and $h_{\omega}^{pp}=h_{\omega}^{(0)}+h_{\omega}^{(1)}$.
Prior to 2001, low energy experiments had constrained only a linear
combination of approximately equal parts $h^{pp}_{\rho}$ and
$h^{pp}_{\omega}$. With the addition of the TRIUMF 221 MeV result in 2001,
$h^{pp}_{\rho}$ and $h^{pp}_{\omega}$ are now separately constrained. The
data so far are not sufficient to determine all 6 couplings, $f_{\pi} ,
h_{\rho}^{0,1,2} , h_{\omega}^{0,1}$, and much careful experimental work
remains to be done. Nonetheless, one should remember that, in the words of
Charles Babbage, ``errors using inadequate data are much less than those
using no data at all''.

\end{document}